\documentclass[aps,showpacs,nofootinbib,superscriptaddress]{revtex4}

\usepackage{graphicx}
\usepackage{bm}
\usepackage{amsmath}
\usepackage{amssymb}
\usepackage{hyperref}

\newcommand{\qslash}{\kern 0.2 em n\kern -0.50em /}
\newcommand{\nslash}{\kern 0.2 em n\kern -0.50em /}
\newcommand{\kslash}{\kern 0.2 em k\kern -0.45em /}
\newcommand{\lslash}{\kern 0.2 em l\kern -0.50em /}
\newcommand{\pslash}{\kern 0.2 em p\kern -0.50em /}
\newcommand{\Sslash}{\kern 0.2 em S\kern -0.50em /}
\newcommand{\Pslash}{\kern 0.2 em P\kern -0.50em /}
\newcommand{\Dslash}{\kern 0.2 em D\kern -0.65em /\kern 0.15em}

\newcommand{\Tr}{\operatorname*{Tr}\nolimits}

\newcommand{\ii}{i}

\usepackage{color}

\begin{document}

\title{Twist-3 fragmentation functions in a spectator model with gluon rescattering}

\author{Zhun Lu}\email{zhunlu@seu.edu.cn}\affiliation{Department of Physics, Southeast University, Nanjing
211189, China}
\author{Ivan Schmidt}\email{ivan.schmidt@usm.cl}\affiliation{Departamento de F\'\i sica, Universidad T\'ecnica Federico Santa Mar\'\i a, and
Centro Cient\'ifico-Tecnol\'ogico de Valpara\'iso,
Casilla 110-V, Valpara\'\i so, Chile}

\begin{abstract}
We study the twist-3 fragmentation functions $H$ and $\tilde{H}$, by applying a spectator model.
In the calculation we consider the effect of the gluon rescattering at one loop level.
We find that in this case the hard-vertex diagram, which gives zero contribution to the Collins function, does contribute to the fragmentation function $H$.
The calculation shows that the twist-3 T-odd fragmentation functions are free of light-cone divergences.
The parameters of the model are fitted from the known parametrization of the unpolarized fragmentation $D_1$ and the Collins function $H_1^\perp$.
We find our result for the favored fragmentation function is consistent with the recent extraction on $H$ and $\tilde{H}$ from pp data.
We also check numerically the equation of motion relation for $H$, $\tilde{H}$ and find that relation holds fairly well in the spectator model.
\end{abstract}


\maketitle

\section{Introduction}

The Collins effect~\cite{Collins:1992kk} has played an important role in the understanding of single spin asymmetries (SSAs)
in various high energy processes, such as semi-inclusive deep inelastic
scattering (SIDIS),
hadron production in pp collision, and
$e^+e^-$ annihilation into hadron pairs.
The mechanism can be traced back to the so called Collins fragmentation function~\cite{Collins:1992kk}, denoted by $H_1^\perp$, which is a transverse momentum dependent (TMD) nonpertubative object entering the factorized description of hard processes.
It originates from the correlation between the transverse momentum of the fragmenting hadron and the transverse spin of the parent quark.
Different from the ordinary unpolarized fragmentation function $D_1$, the Collins function is time-reversal-odd and chiral-odd.
The extraction of the Collins function has been performed in Ref.~\cite{Anselmino:2013vqa}, and in Ref.~\cite{Kang:2014zza} by considering TMD evolution.

For quite some time it was believed that the dominant contribution to the transverse SSA
for hadron production in pp collision comes from the the Qiu-Sterman function $T_F(x,x)$~\cite{Qiu:1991wg,Qiu:1998ia}, which can be related to the transverse-momentum dependent (TMD) Sivers parton density $f_1^{\perp}(x,p_T^2)$~\cite{Sivers:1989cc}: $T_F(x,x) = -\int d^2 p^2_T {p_T^2\over M} f_{1T}^{\perp}(x,p_T^2)|_{\textrm{SIDIS}}$.
The later one also contributes to the Sivers SSA in semi-inclusive deep inelastic scattering (SIDIS) under the TMD factorization.
However, a recent study~\cite{Kang:2011hk} showed that the function $T_F(x,x)$ extracted from $p^\uparrow p \rightarrow hX$ does not match the sign of the Sivers function fitted from SIDIS data.
This is the so called ``sign-mismatch" puzzle.
It was suggested~\cite{Metz:2012ct} that the twist-3 fragmentation contribution may be important for the SSA in pp collision, and could be used to solve the puzzle.
This was further confirmed by a phenomenological analysis~\cite{Kanazawa:2014dca} on SSA of inclusive pion production in $pp$ collision~\cite{Adams:2003fx,Abelev:2008af,Adamczyk:2012xd,Lee:2007zzh} within the collinear factorization, showing that the fragmentation contribution combined with the $T_F(x,x)$ extracted from SIDIS data can well describe the SSAs in $p^\uparrow p \rightarrow \pi X$.
In this framework, three twist-3 fragmentation functions, $\hat H(z)$, $H(z)$ and $\hat H_{FU}^{\Im}(z,z_1)$, participate.
The first one corresponds to the first moment of the TMD Collins function and has been applied to interpret the SSA in $pp$ collisions in previous studies~\cite{Yuan:2009dw,Kang:2010zzb}.
The second one appears in subleading order of a $1/Q$ expansion of the quark-quark correlator, while its TMD version $H(z,\bm k_T^2)$ is also a twist-3 function.
The function $\hat H_{FU}^{\Im}(z,z_1)$ is the imaginary part of $H_{FU}(z,z_1)$, which involves the F-type multiparton correlation~\cite{Yuan:2009dw,Kang:2010zzb,Metz:2012ct}.
The three functions are not independent, as they are connected by the equation of motion relation
\begin{align}
H(z)=-2z\hat H(z)+ 2z^3 \int_z^\infty {dz_1\over z_1^2} \textrm{PV} {1\over {1\over z} -{1\over z_1} } \hat H_{FU}^{\Im}(z,z_1) =-2z\hat H(z)+\tilde{H}(z)\,. \label{eq:eom}
\end{align}
In the last equation we have used $\tilde{H}(z)$ to denote the ``moment" of $H_{FU}^{\Im}(z,z_1)$.
The function $\tilde{H}$ might also contribute to the $\sin\phi_S$ SSA in SIDIS through the coupling with the transversity distribution~\cite{Bacchetta:2006tn}.

Except for $\hat H$, currently the quantitative knowledge about the other twist-3 fragmentation functions mainly relies on the parametrization in Ref.~\cite{Kanazawa:2014dca}.
These fragmentation functions not only play crucial role in the understanding of the SSA in  $pp^\uparrow \rightarrow hX$  process, but also give significant contribution to the SSAs in single-inclusive leptoproduction of hadrons: $\ell p^\uparrow \rightarrow hX$ collision~\cite{Gamberg:2014eia}.
The fragmentation contribution at the twist-3 level also enter the description of the longitudinal-transverse spin asymmetry~\cite{Kanazawa:2014tda}  in the process $\ell^\rightarrow N^\uparrow \to h X$.
Therefore, it is important to perform further theoretical and model study to provide information of $H$ and $\tilde{H}$ complementary to the phenomenological analysis.
Besides, the function $\tilde{H}(z)$ also encodes interesting information regarding the quark-gluon-quark correlation during the parton fragmentation.
In this work we will study those fragmentation functions from the model aspect.
Particularly, we will perform a calculation on the function $H$ and $\tilde{H}$ for the first time,
using a spectator model.
This model has been applied to calculate the Collins function for pions\cite{Bacchetta:2001di,Bacchetta:2002tk,Gamberg:2003eg,Bacchetta:2003xn,Amrath:2005gv,Bacchetta:2007wc} and kaons~\cite{Bacchetta:2007wc}, by considering the pion loop, or the gluon loop.
In our calculation we will incorporate the effect of the gluon loop.
We first calculate the TMD function $H(z,\bm k_T^2)$ and $\tilde{H}(z,\bm k_T^2)$. The corresponding collinear functions are obtained by integrating over the transverse momentum.

\section{Spectator model Calculation of $H$ and $\tilde{H}$}

Here we setup the notations adopted in our calculation.
We use $k$ and $P_h$ to denote the momenta of the parent quark and the final hadron, respectively.
We also apply the following kinematics:
\begin{align}
&k=(k^-, k^+,  \bm k_T) = \left(k^-, {k^2+\bm k_T^2 \over 2k^-}, \bm k_T\right), ~~~P_h=( P_h^-, P_h^+,  \bm 0_T) = \left(zk^-, {M_h^2\over 2zk^-},  \bm 0_T, \right)\,,
\end{align}
where the light-front coordinates $a^\mp = a\cdot n^\pm $ have been used,  $\bm k_T$ denotes the momentum component of the quark transverse to the two light-like vectors $n^\pm$, and $z=P_h^-/ k^-$ is the momentum fraction of the hadron.
The transverse momentum of the hadron with respect to the parent quark direction is given by $\bm K_T = -z \bm k_T$.

\subsection{Calculation of $H$ up to one gluon loop}

\begin{figure*}
  \includegraphics[width=0.24\columnwidth]{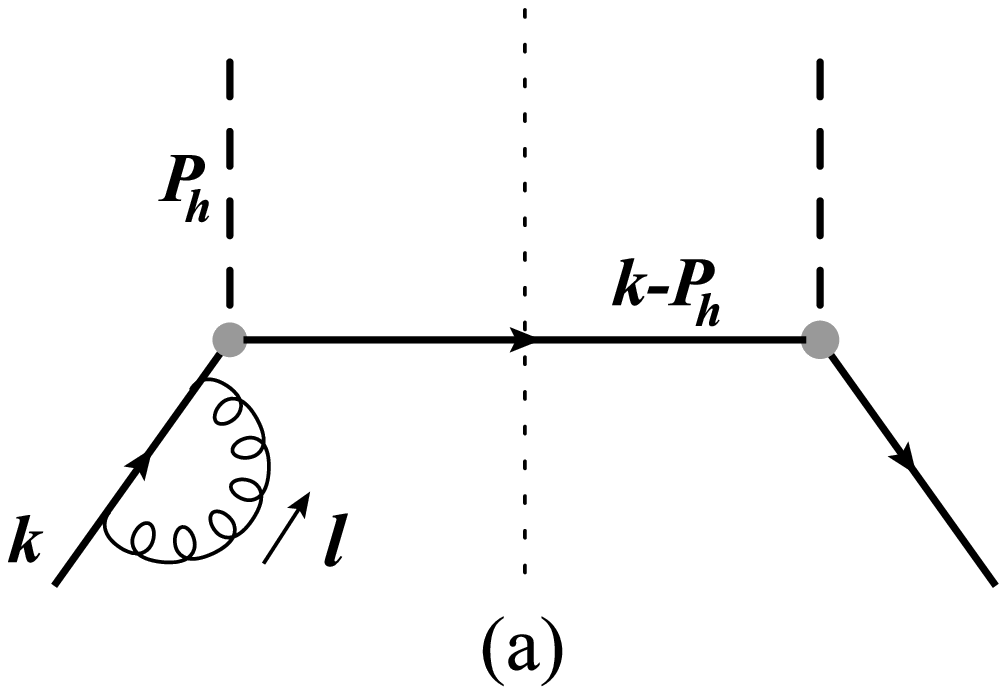}~~~
  \includegraphics[width=0.24\columnwidth]{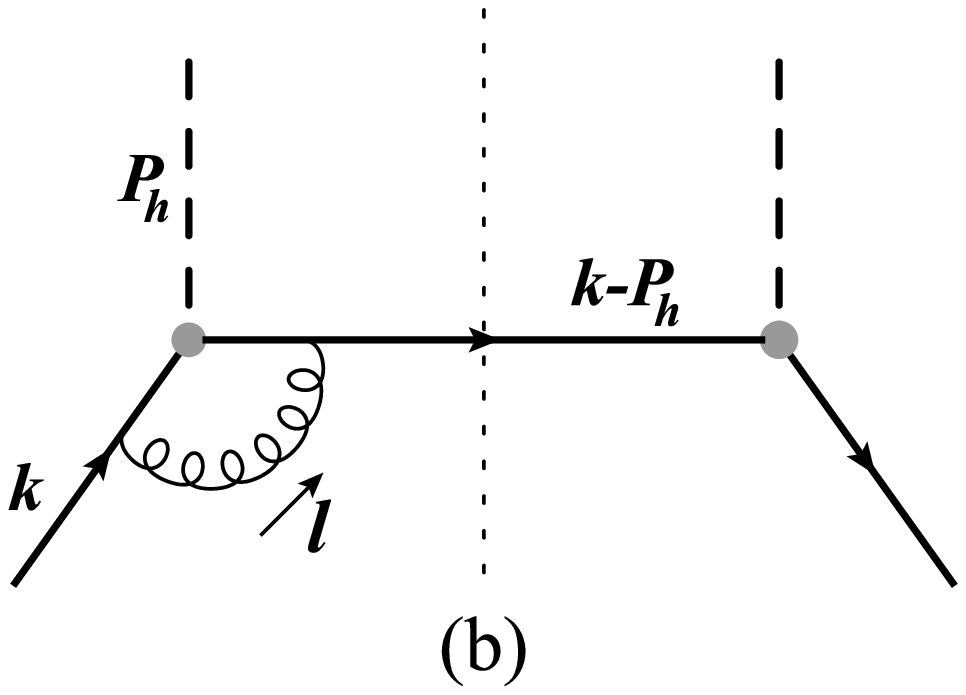}~~~
  \includegraphics[width=0.24\columnwidth]{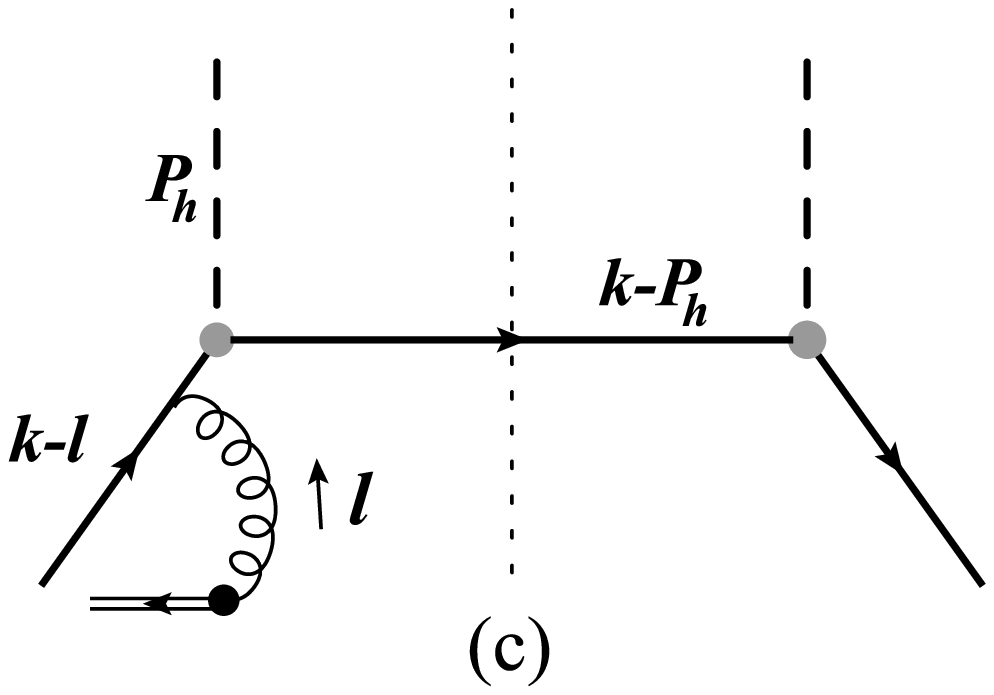}~~~
  \includegraphics[width=0.24\columnwidth]{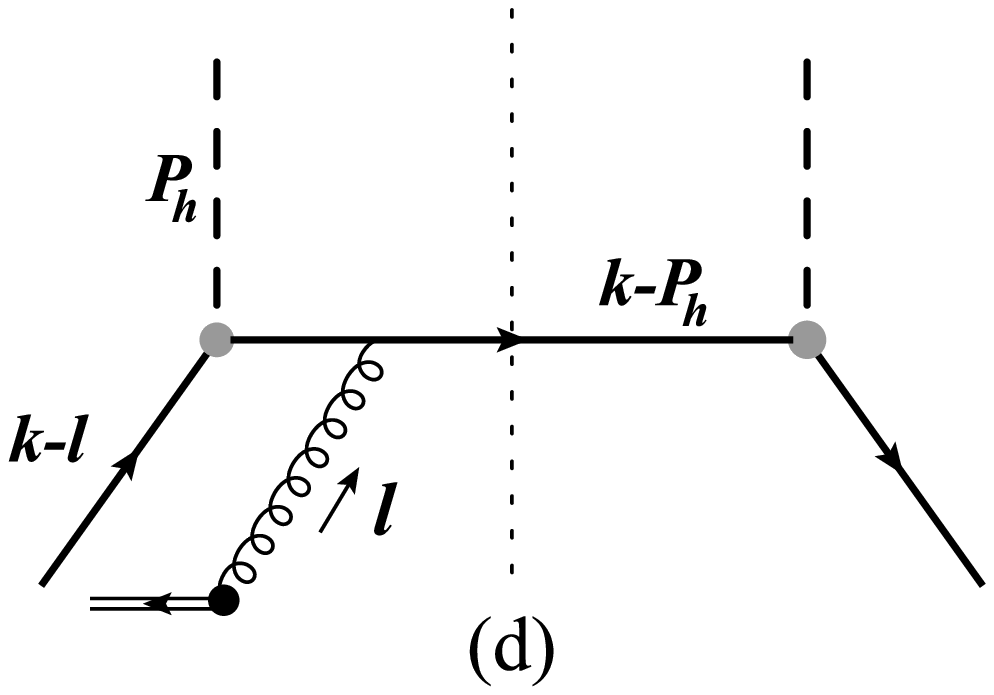}
 \caption {One loop level diagrams utilized to calculate the correlator in the spectator model.
The double lines in (c) and (d) represent the eikonal lines. The hermitian conjugations of these diagrams, which we have not shown here, also contribute.}
 \label{deltadiag}
\end{figure*}

The fragmentation function $H(z,\bm k_T^2)$ can be obtained from the following trace
\begin{align}
{M_h\over P_h^-}\epsilon_T^{\alpha\beta} H(z,\bm k_T^2) = {1\over 2}\textrm{Tr}[\Delta(z,k_T) i\sigma^{\alpha\beta}\gamma_5]\,,
\end{align}
where $\Delta(z,k_T)$ is the TMD correlation function that is defined as:
\begin{multline}
\Delta(z,k_T)  =\frac{1}{2z}\sum_X \, \int
  \frac{d\xi^+  d^2\bm{\xi}_T}{(2\pi)^{3}}\; e^{i k \cdot \xi}\,
    \langle 0|\, {\cal U}^{\infty^+}_{(\bm{\infty}_T,\bm\xi_T)} {\cal U}^{\bm\xi_T}_{({\infty}^+,\xi^+)}
\,\psi(\xi)|h, X\rangle
\langle h, X|
             \bar{\psi}(0)\,
{\cal U}^{\bm{0}_T}_{(0^+,{\infty}^+)} {\cal U}^{\infty^+}_{(\bm{0}_T,\bm{\infty}_T)}
|0\rangle \bigg|_{\xi^-=0}\,.
\label{eq:delta}
\end{multline}
Here ${\cal U}^{c}_{(a,b)}$ denotes the Wilson line running from $a$ to $b$ at the fixed position $c$, to ensure the gauge invariance of the operator.
In the spectator model, the tree level diagrams lead to a vanishing result because of lack of the imaginary phase.
To obtain a nonzero result one has to go to the loop diagrams.
In one-loop level there are four different diagrams (and their hermitian conjugates) that may contribute to the correlator $\Delta(z,k_T^2)$, as shown in Fig.~\ref{deltadiag}.
These include the self-energy diagram (Fig.~\ref{deltadiag}a), the vertex diagram (Fig.~\ref{deltadiag}b), the hard vertex diagram (Fig.~\ref{deltadiag}c), and the box diagram (Fig.~\ref{deltadiag}d).
They have also been applied to calculate the Collins function in Refs.~\cite{Amrath:2005gv,Bacchetta:2007wc}.

We will focus on the the favored fragmentation function, i.e. the fragmentation of $u\rightarrow \pi^+$.
In this case the expressions for each diagram in Fig.~\ref{deltadiag} are as follows:
\begin{align}
\begin{split}
\Delta_{(a)}(z,k_T) &=\ii\frac{4C_F\alpha_s }{2(2 \pi)^2(1-z)P_h^-}\, \frac{(\kslash + m)}{(k^2 -
  m^2)^3}\,  g_{qh}\gamma_5
\,(\kslash -\Pslash_h +m_s) g_{q h}  \gamma_5(\kslash +m)\\
&\quad\int\frac{d^4 l}{(2 \pi)^4} \,
\frac{ \gamma^\mu \,
(\kslash -
  \lslash +m)\,\gamma_\mu\,(\kslash +m) }{((k-l)^2 - m^2 +\ii \varepsilon)(l^2  +\ii \varepsilon)}\,,
\end{split} \displaybreak[0]\\
 \begin{split}
\Delta_{(b)}(z,k_T) &= \ii\frac{4C_F\alpha_s }{2(2 \pi)^2(1-z)P_h^-}\, \frac{(\kslash + m)}{(k^2 -
  m^2)^2}\,  g_{qh}\gamma_5
\,(\kslash -\Pslash_h +m_s)\\
&\quad\int\frac{d^4 l}{(2 \pi)^4} \,
\frac{ \gamma^\mu (\kslash - \Pslash_h - \lslash +m_s)\,
g_{q h}  \gamma_5\,
(\kslash -
  \lslash +m)\,\gamma_\mu\,((\kslash +m) }{((k-P_h-l)^2 -
  m_s^2+\ii \varepsilon ) ((k-l)^2 - m^2 +\ii \varepsilon) (l^2  +\ii \varepsilon)},
\end{split}\displaybreak[0] \,\label{eq:deltab}\\
 \begin{split}
\Delta_{(c)}(z,k_T) &=\ii\frac{4C_F\alpha_s }{2(2 \pi)^2(1-z)P_h^-}\, \frac{(\kslash + m)}{k^2 -
  m^2}\,  g_{qh}\gamma_5
\,(\kslash -\Pslash_h +m_s) g_{q h}  \gamma_5(\kslash +m)\\
&\quad\int\frac{d^4 l}{(2 \pi)^4} \,
\frac{ \gamma^+ \,
(\kslash -
  \lslash +m)\, }{((k-l)^2 - m^2 +\ii \varepsilon) (-l^- \pm i
  \varepsilon)(l^2  +\ii \varepsilon)}\,,
\end{split} \displaybreak[0] \label{eq:deltac}\\
 \begin{split}
\Delta_{(d)}(z,k_T) &= \ii\frac{4C_F\alpha_s }{2(2 \pi)^2(1-z)P_h^-}\, \frac{(\kslash + m)}{k^2 -
  m^2}\,  g_{qh}\gamma_5
\,(\kslash -\Pslash_h +m_s)\\
&\quad\int\frac{d^4 l}{(2 \pi)^4} \,
\frac{ \gamma^+ (\kslash - \Pslash_h - \lslash +m_s)\,
g_{q h}  \gamma_5\,
(\kslash -
  \lslash +m)\, }{((k-P_h-l)^2 -
  m_s^2+\ii \varepsilon ) ((k-l)^2 - m^2 +\ii \varepsilon) (-l^- \pm i
  \varepsilon)(l^2  +\ii \varepsilon)}\,.
\end{split}\label{eq:deltad}
\end{align}
Here $g_{qh}$ is the coupling of the quark-hadron vertex, $m$ the mass of the quark in the initial state, and $m_s$ the mass of the spectator quark.
In Eqs.~(\ref{eq:deltac}) and (\ref{eq:deltad}) we have applied the Feynman rules for the eikonal lines.

In the calculation of T-odd functions, one should utilize the Cutkosky cut rules to put certain internal lines on the mass shell to obtain the necessary imaginary phase.
For T-odd fragmentation functions, only the cuts through the gluon line and the intermediate quark line inside the loop give rise to the result.
This corresponds the following replacements
\begin{align}
{1\over l^2 + \ii\varepsilon} \rightarrow -2\pi i\delta(l^2),~~~~~~~ {1\over (k-l)^2 + \ii\varepsilon} \rightarrow -2\pi i\delta((k-l)^2) \,.\label{eq:cuts}
\end{align}
Here the cuts through the eikonal lines do not contribute.
This directly links to the universality of the TMD fragmentation functions~\cite{
Metz:2002iz,Collins:2004nx,Yuan:2007nd,Gamberg:2008yt}, which has been verified intensively in literature.
Another issue that should be addressed is the choice of the quark-hadron coupling $g_{qh}$.
When choosing the point-like coupling, there is a divergence appearing at large $k_T$ region in the calculation of the collinear fragmentation function:
\begin{align}
H(z) = \int d^2\bm K_T  H(z,\bm k_T^2) =z^2 \int d^2\bm k_T H(z,\bm k_T^2)\,.
\end{align}
In the literature two different approaches have been applied to regularize this divergence.
One strategy is to adopt a cut on $k_T$ by putting an upper limit $k_T^{max}$,
The other is to choose a form factor for $g_{qh}$ which depends on the quark momentum.
Here we will utilize the second approach.
Follow the choice in Ref.~\cite{Bacchetta:2007wc}, we adopt a Gaussian form factor for the coupling,
\begin{align}
g_{qh} \rightarrow g_{qh} {e^{-{k^2\over\Lambda^2}}\over z}\,\label{eq:gaussian}
\end{align}
where $\Lambda^2$ has the general form $\Lambda^2= \lambda^2/(z^\alpha(1-z)^\beta)$.
The $\lambda$, $\alpha$, and $\beta$ are the parameters of the form factor that will be determined in the next section.
The advantage of the choice in Eq.~\ref{eq:gaussian} is that it can also reasonably reproduce~\cite{Bacchetta:2007wc} the unpolarized fragmentation function.

In Eqs.~(\ref{eq:deltab}) or (\ref{eq:deltad}), in principle one of the form factors should depend on the loop momentum $l$.
Here we will drop this dependence and merely use $k^2$ instead of $(k-l)^2$ in that form factor to simplify the integration.
The same choice has also been adopt to calculate the Collins function~\cite{Bacchetta:2007wc}, which is a leading-twist fragmentation function.
For the subleading-twist T-odd functions the situation is more involved.
As shown in Ref.~\cite{Gamberg:2006ru}, the calculation of T-odd twist-3 TMD distributions suffers from a light-cone divergence.
In phenomenological studies the divergence has to be regularized~\cite{Gamberg:2006ru,Lu:2012gu} by introducing form factors, explicitly depending on loop momentum.
However, as we will show later, we find that in the case of twist-3 fragmentation functions, the calculation is free of this light-cone divergence.
The reason behind this distinction is that the kinematical configuration contributing to T-odd fragmentation functions is different from that to the T-odd distribution functions.

After performing the integration over $l$ using the cuts in Eq.~\ref{eq:cuts}, we organize the expression for  $H (z,k_T^2)$ as follows
\begin{align}
H (z,k_T^2) & =   {2\alpha_s g_{q\pi}^2C_F\over (2\pi)^4 }{e^{-2k^2\over \Lambda^2}\over z^2(1-z)}{1\over M_h (k^2-m^2)}\left({H}_{(a)} (z,k_T^2) +{H}_{(b)} (z,k_T^2) +{H}_{(c)}(z,k_T^2)+{H}_{(d)} (z,k_T^2)\right)\,. \label{eq:hzkt}
\end{align}
The four terms in the bracket of the right hand side of (\ref{eq:hzkt}) have the forms
\begin{align}
{H}_{(a)} (z,k_T^2)
 &= -{m\over 2(k^2-m^2) }(3-{m^2\over k^2}) (k^2-m_s^2 + (1-2/z)m_h^2) I_{1}\,, \\
{H}_{(b)} (z,k_T^2) &= \left({k^2-m_h^2+m_s^2\over \lambda(m_h,m_s)}I_{1}-m_s I_{2}\right)(k^2-m_s^2 + (1-2/z)m_h^2)\,, \\
{H}_{(c)} (z,k_T^2) &= -((m_s-m)(k^2-mm_s)+mm_h)I_{1}/(k^2-m^2) - (m_s-m+zm)I_{3}k^-\,,
 \label{eq:thc}\\
{H}_{(d)} (z,k_T^2) &=  {I_{2}\over 2z k_T^2}\bigg((m_s-m+zm)\big(\lambda(m_s,m_h)+\left((1-2z)k^2+m_h^2-m_s^2\right)\left(k^2-m_s^2+(1-2/z)m_h^2\right)\big)\bigg)\nonumber\\
&-zm
\left(k^2-m_s^2+(1-2/z)m_h^2)\right)I_{2} -I_{2}\left((m_s-m)(k^2-mm_s)+mm_h^2\right)+
 (m_s-m+zm)I_{3}k^-\,.
\end{align}
The functions $I_{i}$ represent the results of the following integrals
\begin{align}
I_{1} &=\int d^4l \delta(l^2) \delta((k-l)^2-m^2) ={\pi\over 2k^2}\left(k^2-m^2\right)\,, \\
I_{2} &= \int d^4l { \delta(l^2) \delta((k-l)^2-m^2)\over (k-P_h-l)^2-m_s^2}
=-{\pi\over 2\lambda(m_h,m_s) }  \ln\left(1+{2\sqrt{ \lambda(m_h,m_s)}\over k^2-m_h^2+m_s^2 + \sqrt{ \lambda(m_h,m_s)}}\right)\,, \\
I_{3} & =  \int d^4l {\delta(l^2) \delta((k-l)^2-m^2) \over -l^- +i\varepsilon}\,,
\end{align}
with $\lambda(m_h,m_s)=(k^2-(m_h+m_s)^2)(k^2-(m_h-m_s)^2)$.

We would like to point out that the quark-photon hard-vertex diagram gives nonzero contribution to $H(z,k_T^2)$, as shown in Eq.~\ref{eq:thc}.
This is different from the calculation of the Collins function $H_1^\perp$, in which case the contribution from the hard-vertex diagram vanishes~\cite{Bacchetta:2007wc}.
We note that this is because the Dirac structure of $H(z,k_T^2)$ appearing in the decomposition of the correlation function $\Delta(z,k_T)$ is different from that of the Collins function.
The sum of ${H}_{(c)}(z,k_T^2)$ and ${H}_{(d)} (z,k_T^2)$ can be cast into
\begin{align}
{H}_{(c+d)} (z,k_T^2) &=
{I_{2}\over 2z k_T^2}\bigg((m_s-m+zm)\big(\lambda(m_s,m_h)+\left((1-2z)k^2+m_h^2-m_s^2\right)\left(k^2-m_s^2+(1-2/z)m_h^2\right)\big)\bigg)\nonumber\\
&-zm
\left(k^2-m_s^2+(1-2/z)m_h^2)\right)I_{2} -
\left({I_{1}\over k^2-m^2}+I_{2}\right)\left((m_s-m)(k^2-mm_s)+mm_h^2\right)\,,
\end{align}
where the terms containing $I_{3}$ cancel out.
As we can see, the final result of $H (z,k_T^2)$ in Eq.~(\ref{eq:hzkt}) is free of the light-cone divergence.

\subsection{Calculation of $\tilde{H}$ with gluon rescattering}

The fragmentation function $\tilde{H}(z,k_T^2)$ originates from the quark-gluon-quark (qgq) correlation~\cite{Pijlman:2006vm,Bacchetta:2006tn}:
\begin{align}
\tilde{\Delta}_A^\alpha(z,k_T) &=\sum_{X}\hspace{-0.55cm}\int \; \frac{1} {2zN_c}\int \frac{d\xi^{+}d^2\bm\xi_T} {(2\pi)^3}\int  e^{\ii k\cdot \xi} \langle 0| \int^{\xi^+}_{\pm\infty^+} d{\eta^+}\mathcal{U}^{\bm\xi_T}_{(\infty^+,\eta^+)}\nonumber\\
 &\times gF^{-\alpha}_\perp (\eta) \mathcal{U}^{\bm\xi_T}_{(\eta^+,\xi^+)} \psi(\xi)|P_{h};X\rangle\langle P_{h};X|\bar{\psi}(0)\mathcal{U}^{\bm0_T}_{(0^+,\infty^+)}\mathcal{U}^{\infty^+}_{(\bm 0_T,\bm\xi_T)}|0\rangle\bigg|_{\begin{subarray}{l}
\eta^+ = \xi^+=0 \\ \eta_T = \xi_T \end{subarray}}\,,
\label{eq:qgq}
\end{align}
where $F^{\mu\nu}$ is the antisymmetric field strength tensor of the gluon.
Using the identity
\begin{align}
&\int^{\xi^+}_{\pm\infty^+} d\eta^+ = \pm\int_{-\infty^+}^{\infty^+}  d\eta^+\,\theta(\pm\xi^+\mp\eta^+) \nonumber\\
=& {i\over 2\pi }\int_{-\infty}^{\infty}  d\eta^+ \int d \left(\frac{1} {z}-\frac{1} {z_1}\right) {e^{-i \left(\frac{1} {z}-\frac{1} {z_1}\right)P_h^-(\xi^+-\eta^+)} \over \left(\frac{1} {z}-\frac{1} {z_1}\right) \mp i\epsilon},  \label{eq:hea}
\end{align}
with $\theta$ is the Heaviside function, we can rewrite the qgq correlator as
\begin{align}
\tilde{\Delta}_A^\alpha(z,k_T) &=\sum_{X}\hspace{-0.55cm}\int \; \frac{1} {2zN_c}\int \frac{d\xi^{+}d^2\bm\xi_T d\eta^+} {(2\pi)^4}\int d \left({1\over z}-{1\over z_1}\right)
 {e^{\ii \left({1\over z}-{1\over z_1}\right) p_h^-\eta^+}\over {1\over z}-{1\over z_1}-\ii\varepsilon}
 e^{\ii {P_h^-\over z_1} \xi^-} e^{-\ii \bm k_T\cdot\bm\xi_T}  \nonumber\\
 &\times\langle 0| igF^{-\alpha}_\perp (\eta)  \psi(\xi)|P_{h};X\rangle\langle P_{h};X|\bar{\psi}(0)|0\rangle\bigg|_{\begin{subarray}{l}
\eta^+ = \xi^+=0 \\ \eta_T = \xi_T \end{subarray}}\,.
\label{eq:delta}
\end{align}
Here we have suppressed the Wilson lines for brevity.
In Eqs. (\ref{eq:hea}) and (\ref{eq:delta}) we use $1/z- 1/z_1$ to denote the momentum fraction (along the minus light-cone direction) of the gluon with respect to the final state hadron,  following the notations Ref.~\cite{Metz:2012ct}. Thus $1/z_1$ gives the momentum fraction of the quark correlated with the gluon.

The fragmentation function $\tilde{H}$ can be extracted from the correlator $\tilde{\Delta}_{A}^\alpha(z,k_T) $ by the following projection:
\begin{align}
{1\over 2}\Tr[\tilde{\Delta}_{A}^\alpha(z,k_T) \sigma_\alpha^{\,\,-}] = \tilde{H}(z,\bm k_T^2) + i \tilde{E}(z,\bm k_T^2)\,. \label{eq:proj}
\end{align}
The integrated fragmentation function $\tilde{H}(z) = z^2\int d^2\bm k_T \tilde{H}(z,k_T^2)$
is related to the collinear twist-3 fragmentation function $H_{FU}^{\Im}(z,z_1)$  by
\begin{align}
\tilde{H}(z) = 2z^3 \int_z^\infty {dz_1\over z_1^2} \textrm{PV} {1\over {1\over z} -{1\over z_1} } \hat H_{FU}^{\Im}(z,z_1)\,,
\end{align}
where $\hat H_{FU}^{\Im}(z,z_1)$ is the imaginary part of $H_{FU}(z,z_1)$ that appears in the decomposition of the F-type collinear correlator~\cite{Kang:2010zzb,Metz:2012ct}
\begin{eqnarray}
&&\sum_{X}\hspace{-0.55cm} \int \; \frac{1} {z}\int \frac{d\xi^{+}} {2\pi}\int \frac{d\eta^{+}} {2\pi} e^{i\frac{P_{h}^{-}} {z_{1}}\xi^{+}} e^{i\left(\frac{1} {z}-\frac{1} {z_1}\right)P_{h}^{-}\eta^{+}} \langle 0|igF_{\perp}^{-\alpha}(\eta^{+})\psi(\xi^{+})|P_{h};X\rangle\langle P_{h};X|\bar{\psi}(0)|0\rangle \nonumber \\
&& \hspace{0.5cm} = M_{h}\left[\epsilon_{\perp}^{\alpha\beta}\,\sigma_{\beta}^{\;\,+}\gamma_{5}\,\hat{H}_{FU}(z,z_{1})\right].
\label{eq:F-typeFF}
\end{eqnarray}

\begin{figure*}
  \includegraphics[width=0.4\columnwidth]{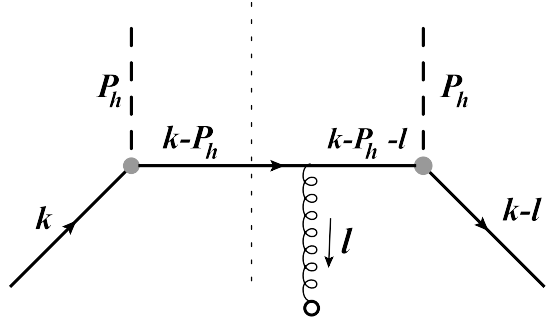}
 \caption {Diagram relevant to the calculation of the qgq correlator in the spectator model}
 \label{fig:qgqspec}
\end{figure*}

The diagram used to calculate the fragmentation function $\tilde{H}$ in the spectator model is shown in Fig.~\ref{fig:qgqspec}, which represents a qgq correlation.
The left hand side of Fig.~\ref{fig:qgqspec} corresponds to the quark-hadron vertex $\langle P_{h};X|\bar{\psi}(0)|0\rangle $, which has the following form in the spectator model
\begin{align}
\bar{U}(P_X) (i\gamma_5) {i(\kslash+m)\over k^2-m^2},
\end{align}
with $P_X$ denoting the momentum of the spectator quark.
The right hand side of Fig.~\ref{fig:qgqspec} corresponds to the vertex $\langle 0|igF_{\perp}^{-\alpha}(\eta^{+})\psi(\xi^{+})|P_{h};X\rangle$, whose expression can be given in a similar way.
The differences are that one should consider the field strength tensor $F^{\alpha\beta}$, as denoted by the circle at the end of the gluon line in Fig.~\ref{fig:qgqspec}.
Its Feynman rule (on the right hand side of the cut) is given by $i(q^\alpha g_T^{\beta\rho} - q^\beta g^{\alpha\rho}) \delta_{ab}$, with $\rho$ and $b$ the indices of the gluon line.
Thus, we can write down the expression for the qgq correlator as:
\begin{align}
 \begin{split}
\tilde{\Delta}_A^\alpha(z,k_T)  &= \ii\frac{4C_F\alpha_s }{2(2 \pi)^2(1-z)P_h^-}\,{1\over k^2-m^2}\\
&\int\frac{d^4 l}{(2 \pi)^4} \,
\frac{(l^-g_T^{\alpha \mu} -l_T^\alpha g^{-\mu}) (\kslash-\lslash + m)\,  g_{qh}\gamma_5
\,(\kslash - \Pslash_h -\lslash +m_s) \gamma_\mu (\kslash - \Pslash_h +m_s)\,
g_{q h}  \gamma_5\,
(\kslash  +m)\, }{(-l^- \pm\ii
  \varepsilon)((k-l)^2 - m^2 -\ii \varepsilon) ((k-P_h-l)^2 -
  m_s^2-\ii \varepsilon ) (l^2  -\ii \varepsilon)}\,, \label{eq:qgqexp}
\end{split}
\end{align}
where we have used the replacement
\begin{align}
 \left({1\over z}-{1\over z_1}\right) P_h^-\rightarrow  l^-\,.
\end{align}

According to Eqs.~(\ref{eq:proj}) and (\ref{eq:qgqexp}), the contribution to $\tilde{H}$ comes from the imaginary part of sub-diagram shown on the right hand side of the cut in Fig.~\ref{fig:qgqspec}.
In order to do this, again one needs to apply the Cutkosky cutting rules to integrate over the internal momentum $l$, that is, to consider all the possible cuts on the propogators appearing in Eq.~(\ref{eq:qgqexp}).
However, only the cuts on the gluon line and the fragmenting quark survive, as shown by the short bars in Fig.~\ref{fig:qgqspec}.
Other combinations of cuts are kinematically forbidden or cancel out each other.
In particular, the total contribution from the pole of the eikonal propagator is zero.
To demonstrate this, we consider two different cases.
The first case is to take the poles of $1/(-l^-\pm \ii\varepsilon)$ and $1/(l^2-\ii\varepsilon)$, therefore, $l_T$ has to be zero.
This yields vanishing $\tilde{H}$ since there is a factor $l^-g_T^{\alpha \mu} -l_T^\alpha g^{-\mu}$ in the numerator of Eq.~(\ref{eq:qgqexp}).
The second case is that one applies the cut on $1/(-l^-\pm \ii\varepsilon)$ and $1/((k-l)^2 - m^2 -\ii \varepsilon)$, or on  $1/(-l^-\pm \ii\varepsilon)$ and $1/((k-P_h-l)^2 -
  m_s^2-\ii \varepsilon )$.
However, these two contributions cancel out each other.
This is because the pole positions for $l^+$ from the propogators $1/((k-l)^2 - m^2 -\ii \varepsilon)$ and $1/((-l^-\pm \ii\varepsilon)$ and $1/((k-P_h-l)^2 -
  m_s^2-\ii \varepsilon )$ are on the same half plane, which means that the integration over $l^+$ vanishes with the delta function $\delta(l^-)$  (since $k^--P_h^- > 0$)
\begin{align}
&\int {dl^+\over 2\pi} {1\over  ((k-l)^2 - m^2 -\ii \varepsilon)  ((k-P_h-l)^2 -
  m_s^2-\ii \varepsilon )} \cdots \nonumber\\
& \sim \int {dl^+\over 2\pi} {1\over  (2k^-(k^+ - l^+) + \cdots  -\ii \varepsilon)  (2(k^--P_h^-)(k^+ - P_h^+ -l^+) +\cdots
 -\ii \varepsilon )} \cdots = 0
\end{align}
Therefore, we will again apply the cutting rules given in Eq.~(\ref{eq:cuts}) to perform the integration over $l$,
and the factor $1/(1/z-1/z_1\pm\ii\varepsilon)$ will take the principal value, as also shown in Refs.~\cite{Liang:2012rb,Metz:2012fq}.
This means that $\tilde H$ is process independent in the spectator model, similar to the Collins function and $H$.
The final result for $\tilde{H}$ has the form
\begin{align}
 \tilde{H}(z,\bm k_T^2)& =
 \frac{\alpha_s g_{q\pi}^2}{(2 \pi)^4}\,C_F\,\frac{e^{-\frac{2k^2}{\Lambda^2}}}{z^2}\,\frac{z}{(1 - z)}\,\frac{1}{M_h(k^2 - m^2)} \bigg{\{}-\mathcal{A}(m_s-m)\bm{k}^2_T \nonumber\\
&+ (m_s-m+zm)\big{[}\mathcal{A}(k^2+\bm k_T^2 ) +\mathcal{B} M_h^2/z  - I_{1}/z-(k^2-m^2)I_{2}/z\big{]}
 \nonumber\\
&+ \big{[}(k^2-mm_s)(m_s-m)+mm_\pi^2\big{]} \big{[}I_{1}/(2zk^2)+\mathcal{A}/z+\mathcal{B}\big{]}\bigg{\}}\,. \label{htilde}
\end{align}
Here $\mathcal{A}$ and $\mathcal{B}$  denote the following functions
\begin{align}
\mathcal{A}&={I_{1}\over \lambda(m_h,m_s)} \left(2k^2 \left(k^2 - m_s^2 - m_h^2\right) {I_{2}\over \pi}+\left(k^2+m_h^2 - m_s^2\right)\right), \\
\mathcal{B}&=-{2k^2 \over \lambda(m_h,m_s) } I_{1}\left (1+{k^2+m_s^2-m_h^2 \over \pi} I_{2}\right),
\end{align}
which appears in the integration
 \begin{align}
\int d^4l { l^\mu\, \delta(l^2)\, \delta((k-l)^2-m^2)\over (k-P_h-l)^2-m_s^2}
=\mathcal{A}\, k^\mu + \mathcal{B}\, P_h^\mu.
\end{align}

\section{numerical result}

In this section we present the numerical result for the fragmentation functions $H$ and $\tilde{H}$.
To this end the values of the parameters in the model have to be specified.
In Ref.~\cite{Bacchetta:2007wc} the parameters of the model were determined by fitting the model result of unpolarized fragmentation function $D_1(z)$ with the Krezter parameterization~\cite{Kretzer:2000yf} of $D_1(z)$.
The parameters were then used to make prediction on the Collins function.
In this paper we will obtain the parameters by fitting simultaneously the model calculations of the unpolarized fragmentation function and the Collins function with the known parameterizations of them, since the Collins functions have been extracted and are well constrained by the $e^+ e^-$ annihilation data and the SIDIS data.
Specifically, we will use the half-$k_T$ moment of the Collins function
\begin{align}
H_1^{\perp (1/2)} (z)= z^2\int d^2\bm k_T {|\bm k_T|\over 2m_h}  H_1^\perp(z,k_T^2)
\end{align}
in the fit.

For the theoretical expressions of $D_1$ and $H_1^\perp$, we use the calculation in the same model, which has already been done in Ref.~\cite{Bacchetta:2007wc}
{\footnote{We recalculate the Collins function and find that our result does not exactly agree with the result in Ref.~\cite{Bacchetta:2007wc}.
For completeness we present our result for $H_1^\perp$ in the Appendix.}}.
For the parameterization of $D_1$, we will adopt the DSS leading order set~\cite{deFlorian:2007aj}.
For the parameterization of the Collins function, we apply the recent extraction By Anselmino et.al.~\cite{Anselmino:2013vqa}.
We note that in Ref.~\cite{Anselmino:2013vqa}, the DSS fragmentation function is also used to extract the Collins function.

Our model calculation is valid at the hadronic scale which is rather low, while the standard parametrization of $D_1$ is usually given at $Q^2>1 \, \textrm{GeV}^2$.
Therefore we extrapolate the DSS $D_1$ fragmentation to that at the model scale $Q^2=0.4\,\textrm{GeV}^2$ in order to perform the fit.
For the same reason, the Collins function should be evolved at that scale for comparison.
However, the evolution of the Collins function is rather complicated~\cite{Yuan:2009dw,Kang:2010xv,Kanazawa:2013uia}.
In the extraction of the Collins function in Ref.~\cite{Anselmino:2013vqa}, the authors used the assumption that the Collins function evolves in the same way of $D_1(z)$.
The same assumption has also used in Ref.~\cite{Bacchetta:2007wc}
For consistency we will use this assumption since in the fit we use the parametrization of Collins function from Ref.~\cite{Anselmino:2013vqa}.

\begin{table}
\begin{tabular}{c|c|c|c|c|c}
  \hline
  ~$m_s$~ (GeV) & $~\lambda$ (GeV)~ &~ $g_{q\pi}~$ &~ $m$ (GeV) &~ $\alpha$ & ~$\beta$  \\
\hline\hline
  0.53 & 2.18 & 5.09 & ~0.3 (fixed)~ &~ 0.5 (fixed)~ &~ 0 (fixed)~ \\
  \hline
\end{tabular}
\caption{Fitted values of the parameters in the spectator model.
The values of the last three parameters are fixed in the fit.}\label{table1}
\end{table}

\begin{figure*}[b]
  \includegraphics[width=0.49\columnwidth]{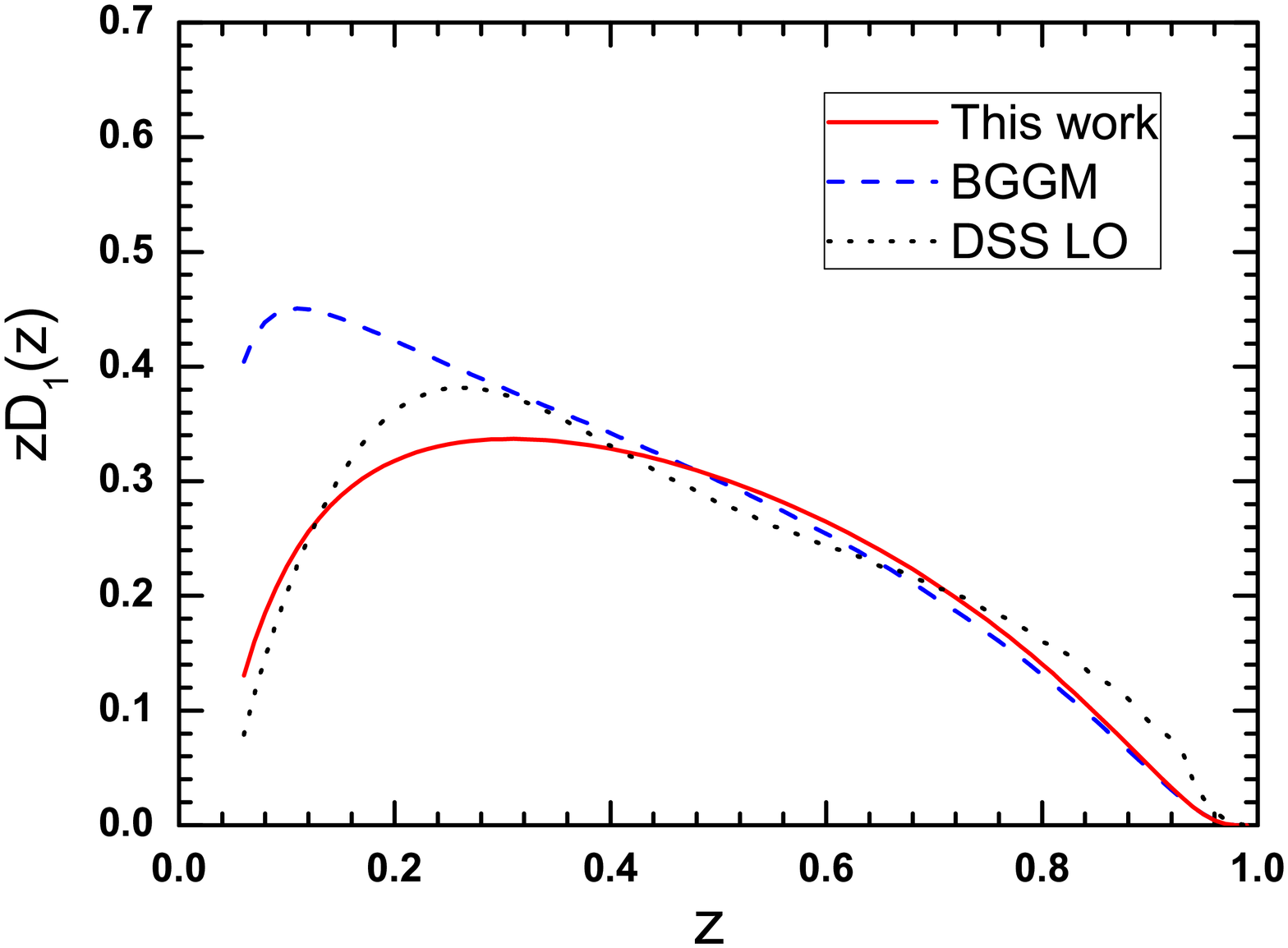}~~
  \includegraphics[width=0.49\columnwidth]{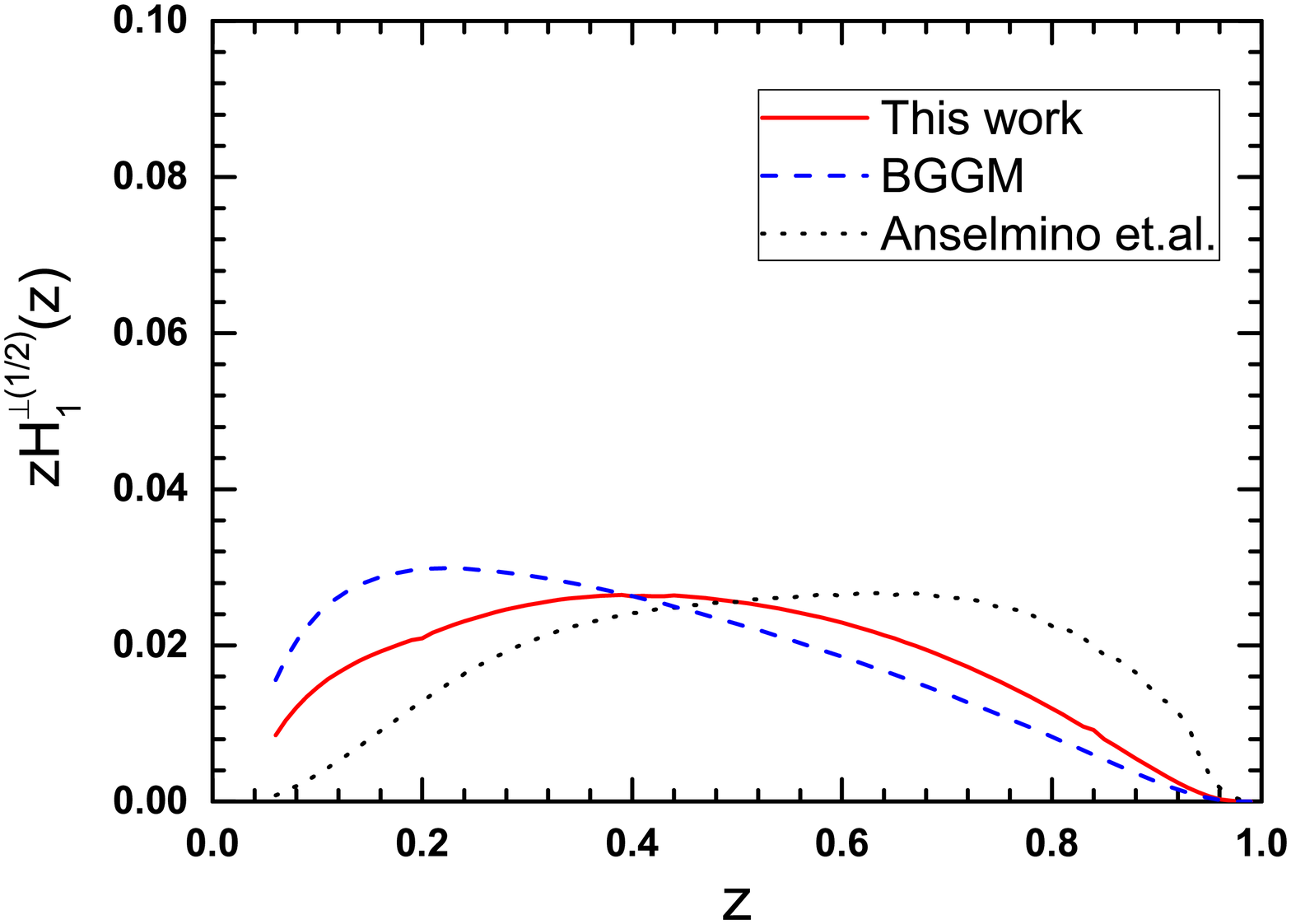}
 \caption { Unpolarized fragmentation function $D_1(z)$ (left panel) and the half moment of the Collins function (right panel) vs $z$ for the fragmentation $u\rightarrow \pi^+$ at the model scale $Q^2=0.4 \textrm{GeV}^2$ .
The parameters are fitted to the parameterizations in Refs.~\cite{deFlorian:2007aj} and \cite{Anselmino:2013vqa}.
The result in Ref.~\cite{Bacchetta:2007wc} (dashed lines) is also shown for comparison.}
 \label{fig:zd1hhalf}
\end{figure*}

In Table.~\ref{table1} we list the fitted values of the parameters in the model.
In the left panal of Fig.~\ref{fig:zd1hhalf}, the curve (the solid line) vs $z$ for the unpolarized fragmentation function $D_1(z)$ at the model scale $Q^2=0.4 \textrm{GeV}^2$ is compared with the curve (dotted line) from the DSS parameterization.
We also show the result (dashed line) calculated from the parameters fitted in Ref.~\cite{Bacchetta:2007wc}.
In the right panal of Fig.~\ref{fig:zd1hhalf}, we display the fitted curve for $H_1^{(1/2)}(z)$ and compare it with the parametrization of Ref.~\cite{Anselmino:2013vqa}.

\begin{figure}
  \includegraphics[width=0.49\columnwidth]{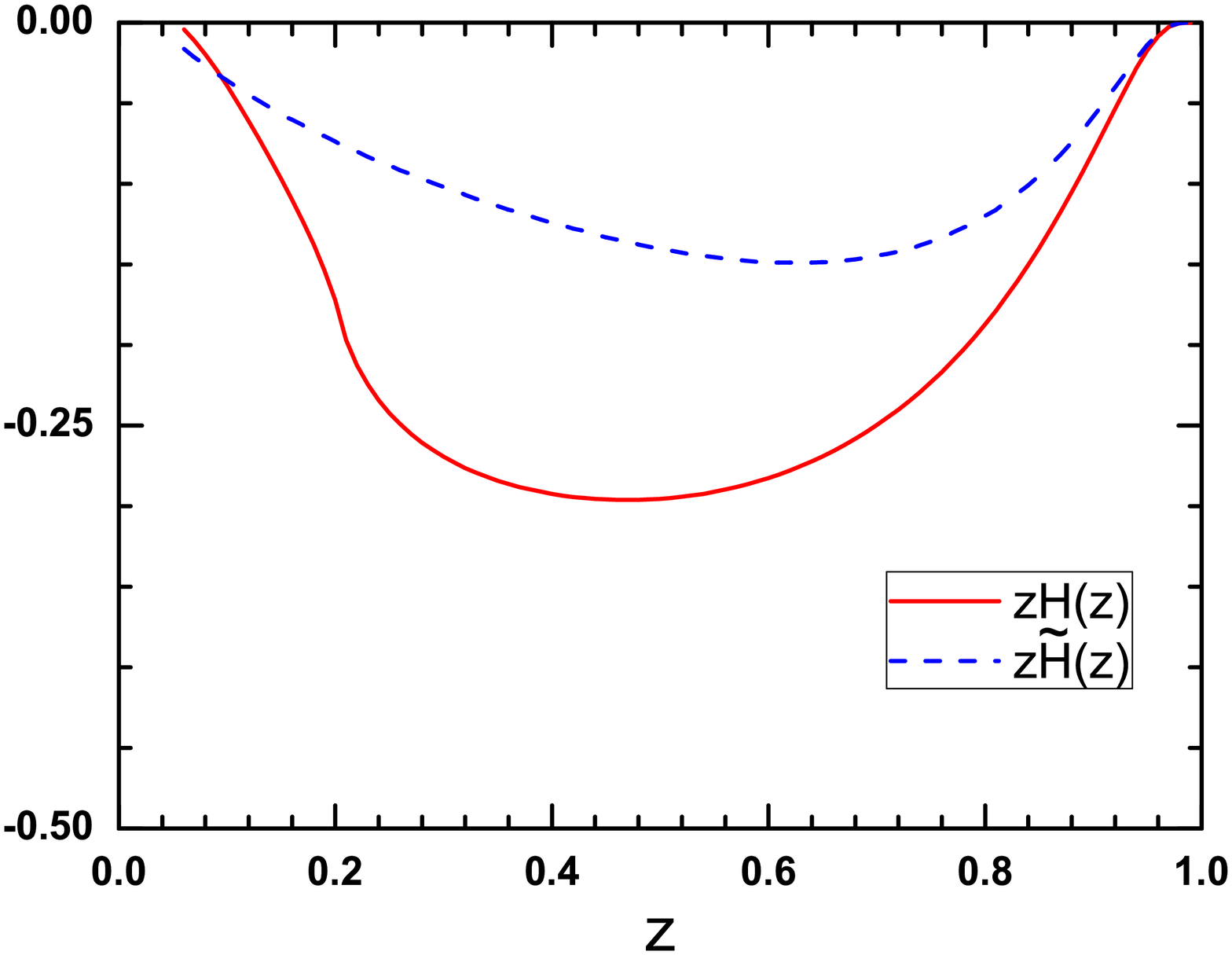}~~
\includegraphics[width=0.49\columnwidth]{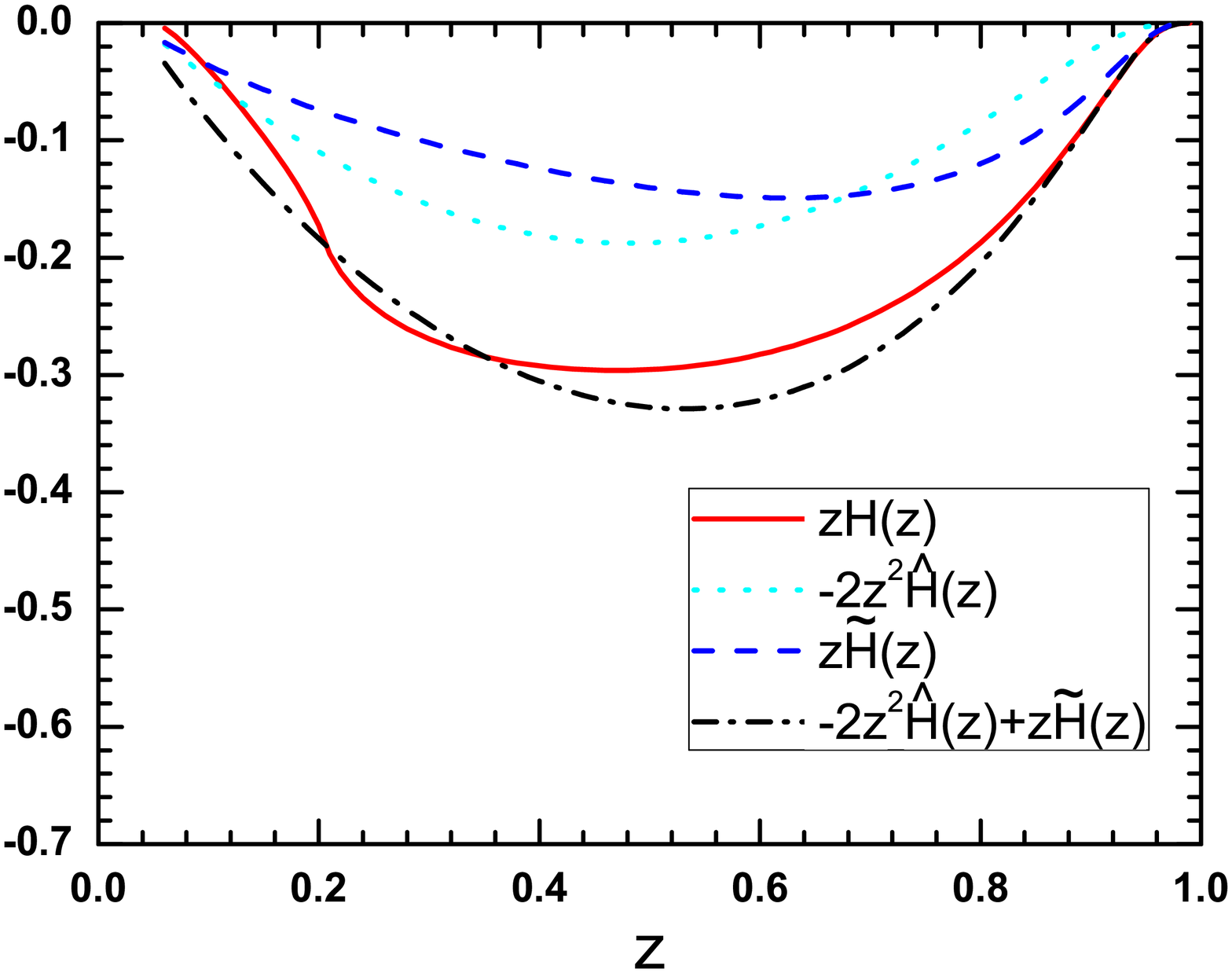}
 \caption {Left panel: The twist-3 fragmentation functions $H(z)$ and $\tilde{H}(z)$ vs $z$, plotted by the solid line and the dashed line, respectively.
Right panel: $H(z)$ compared with $-2z\hat H(z) + \tilde{H}(z)$ in the spectator model.  }
 \label{fig:zhz}
\end{figure}

In the left panel of Fig.~\ref{fig:zhz} we plot our prediction on $H(z)$ and $\tilde H(z)$ using the parameters in Table.~\ref{table1}.
We present the result at the model scale $Q^2=0.4 \textrm{GeV}^2$,
We find that the sign of the favored $H(z)$ is negative and its magnitude is sizable.
This is consistent with the extraction in Ref.~\cite{Kanazawa:2014dca}, where a negative
$H(z)$ for the favored fragmentation is given.
For the function $\tilde H(z)$, we find that the result is nonzero and has a minus sign.
in Ref.~\cite{Kanazawa:2014dca}, a similar result is also hinted by the fit on $H_{FU}^{\Im}(z,z_1)$, which
contribute substantially to $\hat H(z)$ through Eq.~\ref{eq:eom}.

According to Eq.~\ref{eq:eom}, the three twist-3 fragmentation function should satisfy the equation of motion relation, which is a model independent result derived from QCD.
However, From Eqs.(\ref{eq:hzkt}), (\ref{htilde}) and (\ref{eq:collins}), one can not find out an obvious relation among them since in the spectator model they are calculated from different diagrams.
Thus we numerically check the relation (\ref{eq:eom}) and show the the comparison between $H(z)$ (solid line) and $-2z\hat H(z)+ \tilde H(z)$ (dashed-dotted line) on the right panel of Fig.~\ref{fig:zhz}.
We find that the two curves are close, which indicates that the relation holds approximately in the model, therefore it  provide a crosscheck on the validity of our calculation.

\section{Conclusion}

In this work, we studied the twist-3 fragmentation function for $H$ and $\tilde H$ in a spectator model.
We first calculated the TMD functions $H(z,\bm k_T^2)$ and $\tilde H(z,\bm k_T^2)$, and then we obtained the corresponding collinear functions by integrating over the transverse momentum.
In our study we considered the gluon rescattering effect and found that the hard-vertex diagram gives nonzero contribution to $H$.
Using the parameters fitted to the known parameterizations of $D_1$ and $H_1^\perp$ simultaneously, we presented numerical results of $H$ and $\tilde H$.
We found that our results agree with the recent extraction from the SSA in pp collision.
We also tested the equation of motion relation among $\hat H(z)$, $H(z)$ and $\tilde H(z)$,
the numeric result shows that the relation approximately holds in our calculation.
Our study may provide useful information on the twist-3 fragmentation function complementary to phenomenological analysis.

\section*{Acknowledgements}
This work is partially supported by the National Natural Science
Foundation of China (Grants No.~11120101004 and No.~11005018), by the Qing Lan Project (China), and by Fondecyt (Chile) grant 1140390.
Z. L. is grateful to the hospitality of Universidad T\'ecnica Federico Santa Mar\'{\i}a  during a  visit.

\appendix

\section*{Appendix A: Results of the Collins function}

Here we present the model result of the Collins function~\cite{Bacchetta:2007wc}
\begin{align}
H_1^\perp (z,k_T^2) & =  - {2\alpha_s g_{q\pi}^2C_F\over (2\pi)^4 }{e^{-2k^2\over \Lambda^2}\over z^2(1-z)}{M_h\over  (k^2-m^2)}\left({H}^\perp_{1(a)} (z,k_T^2) +{H}^\perp_{1(b)} (z,k_T^2) +{H}^\perp_{1(d)} (z,k_T^2)\right) \label{eq:collins}
\end{align}
The three terms in the brackets correspond to the results from Fig.~\ref{deltadiag}a, Fig.~\ref{deltadiag}b, and Fig.~\ref{deltadiag}d, respectively.
In our calculation we find that those terms have the form
\begin{align}
{H}^\perp_{1(a)} (z,k_T^2)
 &= {m\over (k^2-m^2) }\left(3-{m^2\over k^2}\right)  I_{1} \\
H_{1(b)}^{\perp} (z,k_T^2) &= 2m_sI_{2} -2(m_s-m) \left({m_\pi^2-m_s^2-k^2 \over \lambda(m_h,m_s)}I_{1} - {4k^2m_s^2\over \lambda(m_h,m_s)\pi} I_{1} I_{2} \right) \label{eq:collinsb}\\
H_{1(d)}^{\perp} (z,k_T^2)& = {1\over 2z \bm k_T^2} \left.\{-I_{34}(2zm+2m_s-2m)
+ I_{2} \left[2zm\left(k^2-m^2+M_h^2(1-2/z)\right)\right.\right. \nonumber\\
& +\left.\left. 2(m_s-m) \left((2z-1)k^2 - M_h^2 +m_s^2 -zm(m+m_s) \right)\right]\right\}\,.\label{eq:collinsd}
\end{align}
Here $I_{34}$ is the combination of two integrals
\begin{align}
I_{34}= k^- \left(I_3 + (1-z)(k^2-m^2)I_4 \right) = \pi\ln\left[{\sqrt{k^2(1-z)}\over m_s}\right]
\end{align}
with
\begin{align}
I_{4} &= \int d^4l { \delta(l^2) \delta((k-l)^2-m^2)\over(-l^- + \ii\varepsilon) (k-p-l)^2-m_s^2}
\end{align}

We find that in (\ref{eq:collinsb}) there is a new term proportional to $m_s-m$ that was not contained in Eq.~(29) of Ref.~\cite{Bacchetta:2007wc}.
Also in Eq.~(\ref{eq:collinsd}) the coefficient of certain terms containing $m_s-m$ has a factor of 2 compared to Eq.~(30) of Ref.~\cite{Bacchetta:2007wc}.
But our calculation returns to the results in Ref.~\cite{Amrath:2005gv} in the case $m_s=m$ and by setting the form factor to 1.

\end{document}